# DESARROLLO DE CAPAS DE AlN SOBRE ALEACIONES DE ALUMINIO COMO PROTECCIÓN SUPERFICIAL CONTRA LA CORROSIÓN Y EL DESGASTE

Bürgi J.(1), García Molleja J.(1), Nosei L.(2), Roudet F.(3), Ferrón J.(4), Feugeas J.(1)

*(1) Instituto de Física Rosario – CONICET – Universidad Nacional de Rosario*
*(2) Instituto de Mecánica Aplicada y Estructuras – Facultad de Ciencias Exactas e Ingeniería – Universidad Nacional de Rosario.*
*(3) Institut Universitaire de Technologie « A » de Lille. Université de Lille I*
*(4) Instituto de Tecnología Química – CONICET – Universidad Nacional del Litoral*

El aluminio y sus aleaciones tratados superficialmente en forma adecuada pueden ofrecer excelentes propiedades mecánicas, tribológicas, eléctricas y químicas (resistencia a la corrosión). Sin embargo normalmente no presentan resistencias importantes al desgaste y sus durezas no superan los 170/180 HB dependiendo de la aleación y el tratamiento térmico. En cuanto a la resistencia a la corrosión por cloruros se pueden considerar buenas a la del Al con bajo contenido de impurezas (serie 1000), las aleaciones de la serie 3000, las 5000 (con menos de 3 % de Mg) y las 6000 especialmente aquellas que contienen Mn. En cambio las correspondientes a la serie 2000, las 5000 (con > 3 % de Mg) y las 7000 (con algo de Cu) presentan una baja resistencia a la corrosión [1]. Existen muchos métodos de tratamiento superficiales basados en procesos mecánicos, químicos, electroquímicos y difusivos, pero también basados en la deposición de materiales simples y compuestos en forma de capas gruesas o delgadas, y eventualmente por combinación de varios de ellos. Recientemente, y con el desarrollo de la física de plasmas y la tecnología de descargas eléctricas y aceleración de partículas, un gran número de nuevos procesos se encuentran a disposición para ser estudiados y aplicados en la tecnología del aluminio. En el tratamiento de superficies de metales y aleaciones cabe mencionar a la nitruración [2] y cementación [3] iónicas, en donde mediante descargas eléctricas luminiscentes en atmósfera gaseosa a bajas presiones (~ 5 mbar) conteniendo $N_2$ o $CH_4$ (para nitruración y cementación respectivamente) es posible poner en la superficie a tratar una alta densidad de nitrógeno o carbono (atómicos o iónicos), produciéndose su difusión al interior del material generando capas de nitruros o carburos de varios micrones. En particular, la nitruración puede ser efectuada sobre aluminio, generando una capa superficial de AlN [4]. Igualmente es posible el desarrollo en superficie de AlN utilizando haces de iones de N acelerados a altas energías (>50 keV) mediante un procesos físico conocido como implantación iónica [5]. Pero también es posible la deposición de filmes finos (nanométricos) y hasta la generación de recubrimientos gruesos de centenares de micrómetros de espesor utilizando tecnologías basadas en la generación de plasmas. Al mismo tiempo, filmes finos de materiales puros (Al, Ti, Zr, etc.) o compuestos ($Al_2O_3$, AlN, TiN, TiAlN, etc.) pueden ser depositados en superficie mediante técnicas como PAPVD (Deposición Física en Fase Vapor Asistida por

Plasmas), PACVD (Deposición Química en Fase Vapor Asistida por Plasmas), RSM (Sputter Magnetrón Reactivo), entre otros procesos que utilizan plasmas fríos [6]. Es posible por otro lado el desarrollo de recubrimientos gruesos de compuestos del tipo óxidos de Al, Cr, Zr, Ti, o carburos de W, Co, Cr, etc., utilizando plasmas térmicos como los sistemas Plasma Torch o de Arco de Plasma Transferido [7]. Con el propósito de obtener en superficie determinadas propiedades físicas y/o químicas, la ingeniería de superficie puede constar de uno o varios procesos en forma secuencial.

En particular, el nitruro de aluminio (AlN), que en su fase hcp (wurtzita) es termodinámicamente estable, tiene propiedades ópticas [8], eléctricas [9], piezoeléctricas [10], frente a la propagación de ondas sonoras [11] y mecánicas [12] que permiten su uso en un amplio rango de aplicaciones. Pero además de ellas, posee propiedades mecánicas, tribológicas y químicas diferenciadas que pueden inducir al estudio de su uso en problemas relacionados por ejemplo con la mecánica. Una forma de aprovechar esas propiedades es mediante su generación en superficies de metales, en particular sobre aluminio. Una técnica es mediante la nitruración (nitruración iónica), pero la capa de alúmina que naturalmente se halla en superficie obstaculiza la penetración del N y su posterior difusión al interior, resultando un proceso dificultoso y con el desarrollo de AlN sin una estructura cristalina definida. Otra forma es desarrollando el compuesto externamente y depositándolo en forma de capa sobre la superficie.

En este trabajo se depositaron en superficie de probetas de aleación 7075 filmes de AlN utilizando la técnica de RSM [13]. Un esquema del proceso de deposición de capas de AlN se halla indicado en la Fig. 1. Las probetas fueron maquinadas con 3 formas diferentes destinadas a las distintas técnicas de caracterización: Difracción de Rayos X rasantes (GAXRD) y Microscopía SEM (discos de 20 mm de diámetro y 3 mm de espesor); Espectroscopia de Electrones Auger (AES) (prisma de 12 x 6 x 2 mm); y Desgaste y Corrosión (cilindros de 10 mm de diámetro y 30 mm de longitud). Todas fueron limpiadas terminando en un baño con metanol. La deposición fue realizada utilizando un sputter magnetrón con un blanco de Al 99,99 ubicado a 30 mm de distancia de la superficie de las probetas, en una atmósfera de $N_2$ (50 %) y Ar (50 %) a $6,7.10^{-3}$ mbar. Previo a la deposición las superficies de las probetas fueron sometidas a sputtering con Ar (4500 V, $1,3.10^{-2}$ mbar, 60 min), proceso que permite una limpieza final y la eliminación (parcial) de la alúmina desarrollada naturalmente en superficie. Los espesores de AlN resultantes fueron controlados cambiando la distancia blanco-sustrato y el tiempo total de deposición.

Resultados de AES: en la Fig. 2 se pueden observar los perfiles de concentración de los elementos presentes en la capa AlN depositada, la interfase y parte del sustrato. El eje horizontal de la figura, donde se indican los tiempos de sputtering (en min) entre determinaciones AUGER [3,5], corresponde a profundidades de observación, y donde 30 min de sputtering se corresponden con aproximadamente 12 nm. Las concentraciones del aluminio, Al (en estado atómico) y Alox (en estado combinado), y del nitrógeno muestran que el Al se halla en forma de AlN en la capa superficial, hasta una profundidad equivalente de ~ 160 nm, seguida de una interfase de ~ 35 nm en donde el Al pasa gradualmente de su estado combinado al estado atómico, alcanzando el 100 % en el sustrato. Las concentraciones de Al en su estado combinado y de N en la capa muestran una buena estequiometría del AlN desarrollado.

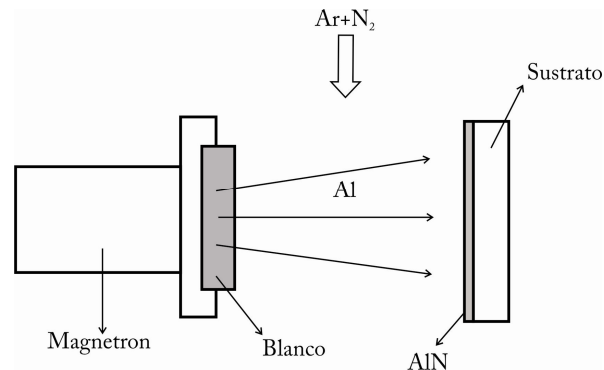

Fig. 1: Esquema del proceso de deposición de filmes de AlN por sputter magnetrón reactivo.

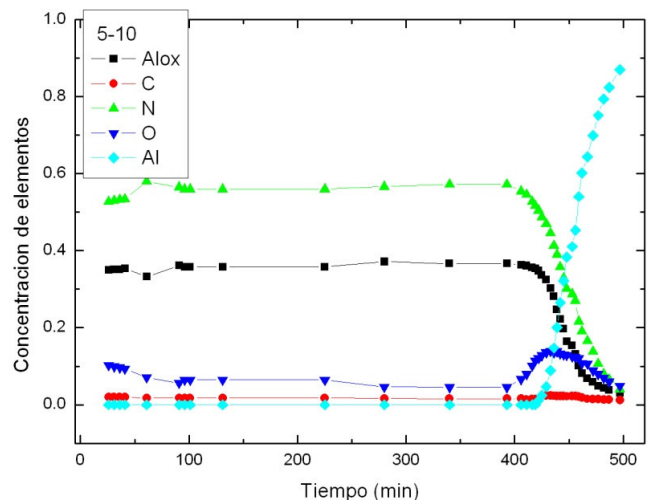

Fig. 2: Perfil de concentración de elementos en superficie por AES. En el eje horizontal, 30 min de sputtering corresponden aproximadamente a 12 nm de remoción de material entre determinaciones AUGER.

Resultados de GAXRD: Los estudios de difracción de rayos X fueron hechos en el modo rasante con un ángulo de incidencia de 1°, lo que permite la observación de las capas más superficiales de las muestras a analizar. En la Fig. 3 se puede observar la existencia de 3 picos correspondientes a los planos (100) en $2\theta=33,1°$, (002) en $2\theta=36,1°$ y (101) en $2\theta=37,9°$ de la estructura hcp (Wurtzitica) del AlN. La existencia de estos 3 picos indica el carácter policristalino de la capa con tendencia a orientaciones según el plano basal.

Resultados de desgaste: los ensayos de desgaste fueron llevados a cabo con un tribómetro Wazau MTM60, consistente en la puesta en contacto de la superficie circular (10 mm de diámetro) de las muestras cilíndricas a estudiar contra un disco de acero AISI 4140 templado y revenido a una dureza de 55 HRc y posteriormente pulido a espejo, de 100 mm de diámetro como puede verse en el esquema de la Fig. 4.

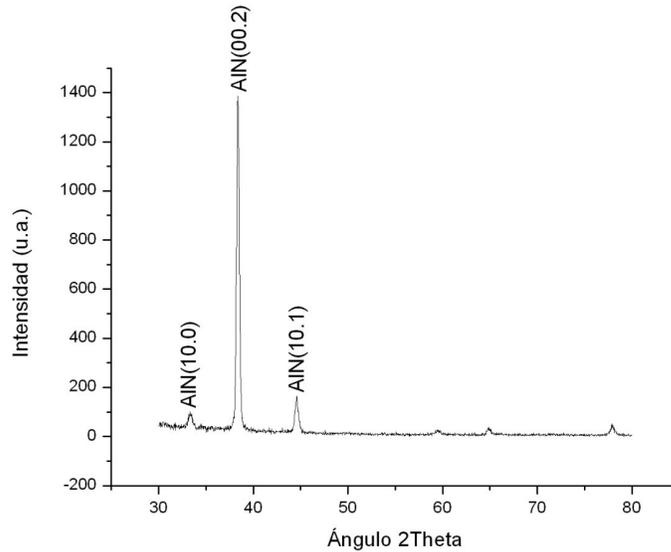

Fig. 3: Difractograma tomado en modo rasante con de incidencia de 1° usando radiación Kα-Cu. Se observan los picos correspondientes a la estructura hcp del AlN.

El contacto es producido con una carga de 10 N actuando desde abajo hacia arriba sobre la probeta ubicada en la parte inferior a 30 mm del centro del disco de acero. El disco, en la parte superior, rota con una velocidad constante de 30 rpm, manteniéndose en ese estado hasta cubrir una distancia de deslizamiento de la probeta sobre él de 100 m, durante 800 s. El desgaste, medido como el acortamiento de la probeta, y la fuerza de fricción son registrados automáticamente en función del tiempo de ensayo. Los resultados, promedio de 4 determinaciones independientes para cada caso (con y sin deposición superficial de AlN), mostraron que mientras las muestras sin recubrir se desgastaron 2,7 µm en promedio, las recubiertas permanecieron sin acortamiento apreciable dentro del margen de apreciación del equipo (±1 µm). Con respecto a las fuerzas de fricción éstas revelaron una reducción de 3,8 N para el caso de las caras sin recubrimiento hasta 2,1 N para el caso de las caras recubiertas con AlN, lo que representa una reducción de ~45 %. Al mismo tiempo las fluctuaciones en la fuerza de fricción en promedio representan 1,3 N y 0,7 N para las caras sin y con el depósito de AlN respectivamente, correspondiendo a una reducción de estas fluctuaciones de ~43 %.

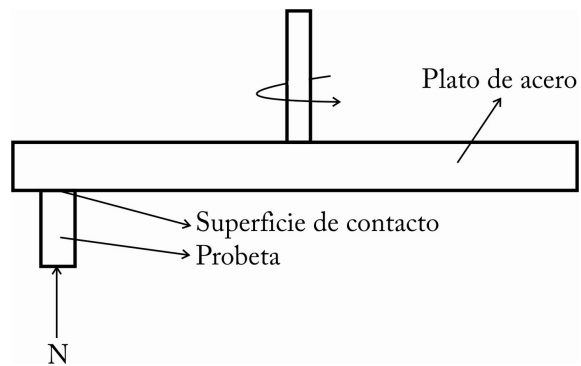

Fig. 4: Esquema del ensayo de desgaste.

Resultados sobre corrosión: Probetas con superficies terminadas según un maquinado con torno sin deposición de AlN en algunos casos y con deposición de AlN en otros fueron expuestas al ataque de cloruros en una cámara de niebla salina con una solución de ClNa al 5 % (p/v), un volumen condensado de 1 cc/h, a 35°C y durante 390 minutos. Al cabo de ese tiempo fueron extraídas y observadas sus superficies por microscopía electrónica de barrido. En la parte a- de la Fig. 5 se puede ver la imagen de la superficie de la aleación de aluminio sin recubrir, mostrando daños severos y productos de corrosión, alcanzando a cubrir regiones enteras de dimensiones de varios centenares de micrones. En la parte b- de la misma figura en cambio, y correspondiendo a una superficie recubierta con AlN, ésta no muestra productos ni daños, apreciándose claramente la textura superficial dejada por el maquinado debido a la transparencia del AlN.

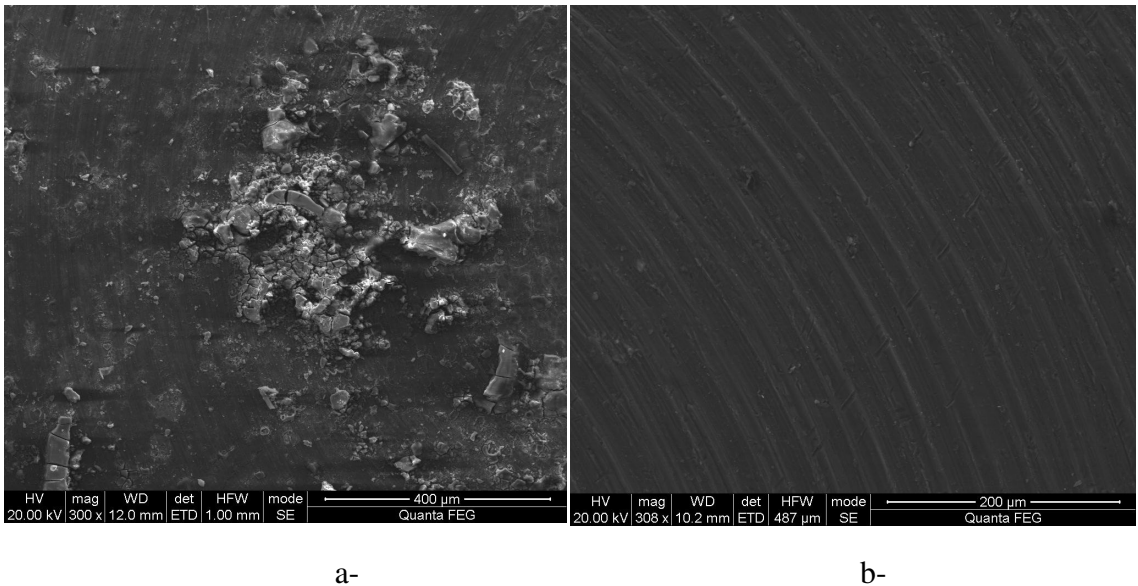

a-                                                                                      b-

Fig. 5: Imágenes de las superficies de aleación 7075 expuestas al ataque de ClNa al 5 % (p/v) en una cámara de niebla (35°C durante 390 segundos). a- Superficie sin recubrimiento de AlN mostrando un severo daño por corrosión, b- superficie recubierta de un film de AlN en la que no se observan daños superficiales.

Conclusiones: Se desarrollaron capas de AlN sobre una aleación de aluminio 7075 utilizando la técnica de Sputter Magnetrón Reactivo. El AlN resultó con estructura wurtzítica policristalina, mostrando que este material en superficie le confiere a esta aleación mejoras en sus propiedades frente al desgaste, reduciendo la fuerza de fricción, y mejorando radicalmente su resistencia a la corrosión provocada por cloruros.

REFERENCIAS